\documentclass[aps,twocolumn,preprintnumbers,amsmath,amssymb,subeqn]{revtex4}

\makeatletter
\def\@eqnnum{{\normalfont \normalcolor [\theequation]}}
\makeatother

\usepackage{amsmath,amssymb,mathrsfs}
\usepackage{graphicx}
\newcommand{\amp}{&\!\!\!\!}
\newcommand{\Bnuc}{{\bf B}_{\mbox{\tiny{nuc}}}}
\newcommand{\BO}{{\bf B}_0} 
\begin{document}

\title{Magnetic Susceptibility: Solutions, Emulsions, and Cells}
\author{P.~W.~KUCHEL$^{1,}$\footnote{Email: p.kuchel@mmb.usyd.edu.au}, B.~E.~CHAPMAN$^1$, W.~A.~BUBB$^1$, P.~E.~HANSEN$^2$, C.~J.~DURRANT$^3$, M.~P.~HERTZBERG$^3$}
\affiliation{$^1$ School of Molecular and Microbial Biosciences, University of Sydney, New South Wales 2006, Australia\\
$^2$ Department of Life Sciences and Chemistry, Roskilde University, 4000 Roskilde, Denmark\\
$^3$ School of Mathematics and Statistics, University of Sydney, New South Wales 2006, Australia}


\begin{abstract}
Differences in magnetic susceptibility between various compartments in
heterogeneous samples can introduce unanticipated complications to NMR spectra. On the
other hand, an understanding of these effects at the level of the underlying physical
principles has led to the development of several experimental techniques that provide
data on cellular function that are unique to NMR spectroscopy. To illustrate some key
features of susceptibility effects we present, among a more general overview, results
obtained with red blood cells and a recently described model system involving diethyl
phthalate in water. This substance forms a relatively stable emulsion in water and yet it has
a significant solubility of $\sim 5$ mmol $L^{-1}$ at room temperature; thus, the NMR spectrum has
twice as many resonances as would be expected for a simple solution. What determines the
relative intensities of the two families of peaks and can their frequencies be manipulated
experimentally in a predictable way? The theory used to interpret the NMR spectra from
the model system and cells was first developed in the context of electrostatics nearly a
century ago, and yet some of its underlying assumptions now warrant closer scrutiny. While
this insight is used in a practical way in this article, the accompanying article deals with the
mathematics and physics behind this new analysis.
\end{abstract}


\maketitle

\section*{INTRODUCTION}

\subsection*{Purpose}
Our aim is to provide an understanding of how differences
in magnetic susceptibility in different regions
of a sample impinge on NMR spectra and magnetic
resonance imaging (MRI) images of cellular systems.
In high-resolution NMR of liquids, magnetic susceptibility
effects are known to impinge on resolution and
the accurate assignment of chemical shifts (e.g., 1),
while in heterogeneous samples the situation is more
complex. Hence, we begin with a brief overview of
the motivation to understand this area as NMR spectroscopists
investigating cells, and then describe the
basic physics of magnetism as it relates to heterogeneous
systems. Following this, we describe an NMR
method for the measurement of magnetic susceptibility
and proceed to show the ``tangible" or visible
spectroscopic effects of changing magnetic susceptibility
in a suspension of red blood cells (RBCs). The
recently studied model system of diethyl phthalate
(DEP) in dilute aqueous emulsion provides an elegant
example of the phenomena arising from differences in
magnetic susceptibility that underlie some contemporary
studies of cellular function in vivo.

\subsection*{Motivation}
It was recently discovered that lipid globules inside
muscle [so-called intramyocellular lipid (IMCL)] and
those globules in subcutaneous and other tissues [socalled
extramyocellular lipid (EMCL)] give $^1$H nuclear
magnetic resonance (NMR, MRS) peaks at different
frequencies in the spectrum (2, 3). This occurs
despite the similarity of the chemical composition of
the lipids in the different tissue compartments. It is
now known that the separate frequencies are simply
the consequence of a combination of the differences
in magnetic susceptibility across the cellular compartments
and the respective shapes of the microcompartments
occupied by the lipids. This is one of the latest
examples of magnetic susceptibility effects being recognized,
understood, and then quantified to make a
new, unique, investigative tool of cellular function in
vivo (4). There has been a lot of work done over many
years on understanding the effect of magnetic susceptibility
differences across tissue compartments on
MRI images (e.g., 5), and of the related effects of
contrast agents (6); and, further insights in this area
have recently been added (7, 8). Inevitably, this work
has rested upon NMR studies of pure chemical systems
(e.g., 1, 9) and a recent review in this journal
addresses fundamental magnetic susceptibility issues
relating to NMR probe design (10).

\subsection*{Spin Echo}
To our knowledge, the first use of differences in
magnetic susceptibility to study an aspect of cellular
function was the exploitation of differential signal
intensity, inside and outside cells, from various solutes
detected in $^1$H spin-echo NMR spectra of RBCs
(erythrocytes) in suspension (11). The physical basis
of the effect is creation of inhomogeneities in the
magnetic field brought about by differences in the
magnetic susceptibility of the cell cytoplasm and the
suspension medium. In the spin-echo experiment
($\pi/2-\tau-\pi-\tau-$acquire) the signal intensity,
$S(2\tau)$, after the echo time $2\tau$, is not only a function of
the intrinsic transverse relaxation time, $T_2$, of the
resonant nuclei but also of the diffusive motion of the
solute molecules that bear the nuclei through inhomogeneous
magnetic fields. Thus,
\begin{eqnarray}
S(2\tau)=S(0)\exp(-2\tau/T_2 - 2\gamma^2g^2D\tau^3/3)F(J)
\end{eqnarray}
where $S(0)$ is the signal intensity when $\tau=0$, $\gamma$ is the
magnetogyric ratio of the nucleus, $g$ is the magnitude
of the magnetic field inhomogeneity expressed to a
first approximation as a linear field gradient, $D$ is the
diffusion coefficient of the solute, and $F(J)$ is a term
that describes the amplitude modulation of the signal
due to spin–spin coupling; it has the value 1 when
there is no coupling (12, 13).

In a suspension of RBCs the average magnetic
field inhomogeneities are larger outside than inside
the cells (11, 14). If the cells were ellipsoidal or
spheroidal and were sufficiently far apart in the suspension
so that they could be considered to be ``isolated,"
the field inside would be uniform while the
field outside would be inhomogeneous (15). These
shapes are surrounded by so called degree-2 surfaces
as they are described by mathematical functions in
which the independent Cartesian variables $x$, $y$, and $z$
are raised to the power 2; a familiar example is a
sphere centred on the origin with a radius $r$, its expression
is $x^2+y^2+z^2=r^2$. Human RBCs are
biconcave discs whose surface can only be described
by at least a degree-4 expression (16). Numerical
solutions of the Laplace equation that yield descriptions
of the magnetic fields in and around an isolated
cell (see below) show that the magnetic field is nonuniform
in both regions.

An early experimentally based consideration of
susceptibility-induced field disturbances in NMR
samples was given by Glasel and Lee (17), who
studied packed beds of glass spheres surrounded by
$^2$H$_2$O. They deduced that there was little bound water
at the surface of the glass beads but that the value of
the apparent $T_2$ was a function of the magnetic field
inhomogeneities at the glass–water boundary. Similarly,
the relative intensity of the extracellular spinecho
signal from oxygenated RBC is reduced outside
compared with inside, when the extracellular magnetic
susceptibility is increased by adding the membrane-
impermeant paramagnetic cage complex Fe(III)–
ferrioxamine to the suspension. In other words, the
specific signal intensity is greater inside the cell as a
result of a longer apparent $T_2$ (defined by the terms in
the exponent in Eq. [1]) that is a consequence of
diffusion of the extracellular solute in the inhomogeneous
magnetic field around the cells. Thus, a membrane-
permeable solute that is transported to the inside
of RBCs in such a suspension, over periods of
minutes to hours, shows progressively increasing
spin-echo signal intensity. It is possible to determine
the transport rate from these time courses, but there
are caveats associated with interpreting these experiments:
Specifically, a change in cell volume, which
can arise if there is not careful control over the osmolality
of added solutions, also alters the magnitude
of field inhomogeneities inside and outside the cells.
Nevertheless, the experiment has been used successfully
to characterize the transport kinetics of alanine,
lactate (11), and choline (18) into human RBCs.

It is evident from the above considerations that a
detailed understanding of magnetic field theory is
required in biomedical NMR spectroscopy. Because
many of the features of magnetic fields in inhomogeneous
media are not described in detail in the bio-NMR 
literature, they are discussed next.

\section*{BASIC CONCEPTS OF MAGNETISM}

\subsection*{Magnetic Field}
A magnetic field is said to exist in a region of space
if a magnet, or a moving electric charge, experiences
a force when placed in it. Lines along which the force
acts throughout the space represent this field graphically.
The representation suggests an analogy with
streamlines in a flowing liquid, and hence it evokes
the concept of magnetic flux (Latin, flow). Thus, the
physical attribute that characterizes a magnetic field is
its flux density (flow per unit area), expressed in units
of joules per ampere per square meter. Rearrangement
of these units, which are called the tesla (T), yields
newtons per meter per ampere (N m$^{-1}$ A$^{-1}$). This
provides one way of visualizing how the magnitude of
the field might be measured: A magnetic field is said
to have a magnetic induction (or flux density) of 1 T
if a conductor of length 1 m, carrying a current of 1 A
and lying at right angles to the flux lines (see Fig. 1),
experiences a force of 1 N.
\begin{figure} [t]
\includegraphics[width=50mm]{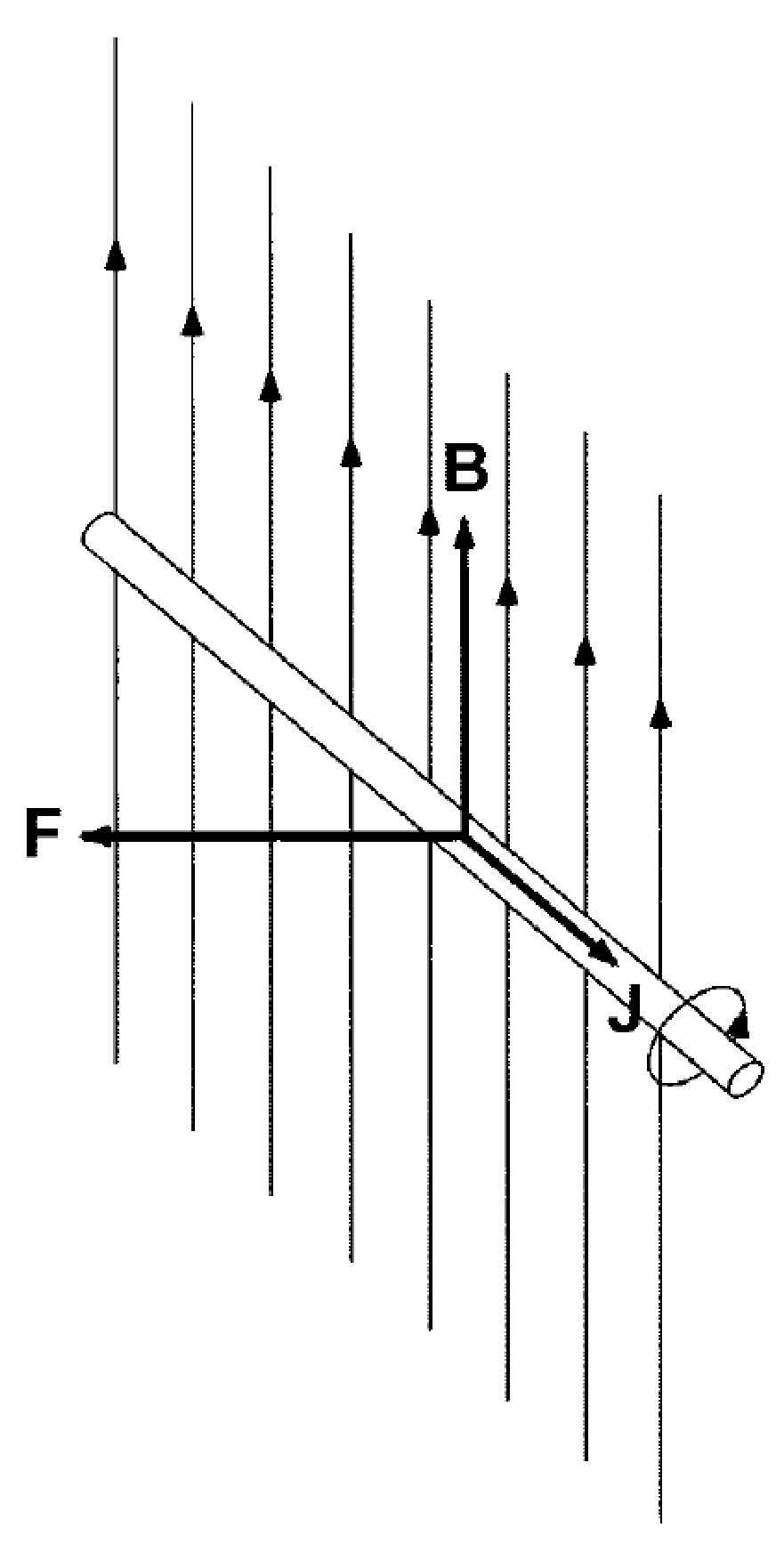}
\caption{Current vector ${\bf J}$ in a conducting wire that is
arranged to be orthogonal to the uniform magnetic field
vector ${\bf B}$; the force ${\bf F}$ results from the interaction between
the two vector fields. Note that the direction of ${\bf F}$ can be
recalled by using Fleming's ``left-hand rule," whereby the
thumb points in the direction of travel (force) of the wire,
the index finger points in the direction of the current (${\bf i}$ or ${\bf J}$),
and the forefinger in the direction of the field (${\bf B}$). Another
way to deduce the direction of force is to note that the
magnetic lines of force around the conductor (see circular
line of force on the right-hand end of the section of wire in
the diagram) are arranged according to the direction of the
fingers when the right hand is wrapped around the wire with
the thumb pointing in the direction of the current. (Note that
the convention is that the direction of the current is that of
the motion of positively charged units; i.e., opposite the
direction of electron flow). And, lines of force pointing in
the same direction repel each other. Hence, in the diagram
the wire would move out of the page as the current-induced
field repels ${\bf B}$ from the rear of the wire.}
\end{figure}

Clearly, magnetic fields exist in matter, and although
``free space" (a vacuum) is free of matter it
also can be the location for a magnetic field. Thus, a
current passing through a conductor of a specified
geometry creates a magnetizing force, called the magnetic
field strength, ${\bf H}$, that, in turn, establishes a
magnetic flux density, ${\bf B}$. 
The actual value of ${\bf B}$ in the
substance will depend on its extent of magnetic polarizability.
Hence, ${\bf H}$ acts on the medium to produce
${\bf B}$, and the simple relationship between these two
properties is
\begin{eqnarray}
{\bf B}=\mu{\bf H}
\end{eqnarray}
where $\mu$ is the magnetic permeability (units, henries
per meter; H m$^{-1}$ = J A$^{-2}$ m$^{-1}$).

In a vacuum the expression has the form
\begin{eqnarray}
{\bf B}=\mu_0{\bf H}
\end{eqnarray}
where $\mu_0$ is the magnetic permeability of free space;
it has the value $4\pi\times 10^{-7}$ H m$^{-1}$. [Note that the
units H (henries) are not to be confused with the
symbol for the magnetizing field.]

The units of ${\bf H}$ are A m$^{-1}$; and one SI unit of
magnetic field strength is defined as that generated at
the center of a circular conductor of diameter 1 m
carrying a current of 1 A. Thus, ${\bf H}$ describes the
physical arrangement of the magnetic field generator
(shape of the current-carrying conductor and its current)
while ${\bf B}$ incorporates this characteristic together
with an expression of the tendency of the medium in
which the field resides to be magnetized.

In a medium, the interaction of the moving charges
in the atoms and molecules within a magnetic field
leads to the induction of a bulk magnetic dipole moment,
denoted by the magnetization ${\bf M}$. It has the units
of magnetic dipole moment (A m$^2$) per unit volume
(m$^{-3}$), or A m$^{-1}$, just like ${\bf H}$. 
Therefore, we write
\begin{eqnarray}
{\bf B}=\mu_0({\bf H}+{\bf M})
\end{eqnarray}
We can derive Eq. [2] from Eq. [4] by introducing a
parameter, $\chi$, called the magnetic susceptibility; it
relates the magnetization of the material to the magnetic
field strength as follows:
\begin{eqnarray}
{\bf M}=\chi{\bf H}
\end{eqnarray}
Thus,
\begin{eqnarray}
{\bf B}&=&\mu_0({\bf H}+\chi{\bf H})\nonumber\\
&=&\mu_0(1+\chi){\bf H}=\mu{\bf H}
\end{eqnarray}
where we note that the permeability of the medium is
given by $\mu=\mu_0(1+\chi)$. The factor
$\mu/\mu_0=(1+\chi)$ is called the relative permeability. 
In diamagnetic materials $\chi<0$ so that 
$\mu<\mu_0$, and in paramagnetic
materials $\chi>0$, so that $\mu>\mu_0$. All materials are
(weakly) diamagnetic but many are also paramagnetic.
For most materials, the paramagnetism is usually
stronger than the diamagnetism at room temperature,
but it decreases with temperature in what is
called the Curie effect (19).

Water and most common gases except oxygen are
diamagnetic. Oxygen (in its low-energy triplet state)
is paramagnetic, as are many ions of the transition
metals.

\subsection*{Larmor Equation}
The master equation of NMR theory is the Larmor
equation; it specifies that the resonance frequency, $\omega$,
of a nucleus is directly proportional to the value of ${\bf B}$
in its immediate neighborhood, viz., 
${\bf\omega}=-\gamma{\bf B}_{\mbox{\tiny{nuc}}}$. In
view of the fact that ${\bf B}_{\mbox{\tiny{nuc}}}$ is a function of the magnetic
susceptibility of the medium, a change in this value
can change the resonance frequency of the nucleus;
but, because the nucleus is surrounded by polarizable
material the field in its immediate vicinity also depends
on the shape of the macroscopic container. In
other words, ${\bf B}_{\mbox{\tiny{nuc}}}$
is also a function of the shape of the
body in which the nucleus resides.

With an understanding of the basis of shifts of
resonances induced by the effects of bulk magnetic
susceptibility (BMS) we can readily predict the {\emph direction}
of shifts in resonance frequency; but see (20, 21)
for caveats. What is more challenging is predicting
the {\emph magnitude} of shifts in variously shaped compartments
of cells and tissues.

\section*{HOW THE GEOMETRY OF THE SAMPLE
AFFECTS ${\bf B}_{\mbox{\tiny{nuc}}}$}

\subsection*{General}
The idea that certain materials when placed in a
magnetic field lead to a distortion of the field is a
familiar one and Fig. 2 shows the nature of such fields
for five bodies of simple geometric form. The actual
calculation of the value of the field at any point in or
around the bodies requires some relatively sophisticated
mathematics and computation, which is outlined
in the next section but is dealt with in detail in the
accompanying article (22). On the other hand, those
wishing to progress rapidly to the more practical
aspects of the general topic of magnetic susceptibility
can safely skip the following subsection.
\begin{figure}[t]
\includegraphics[height=134mm]{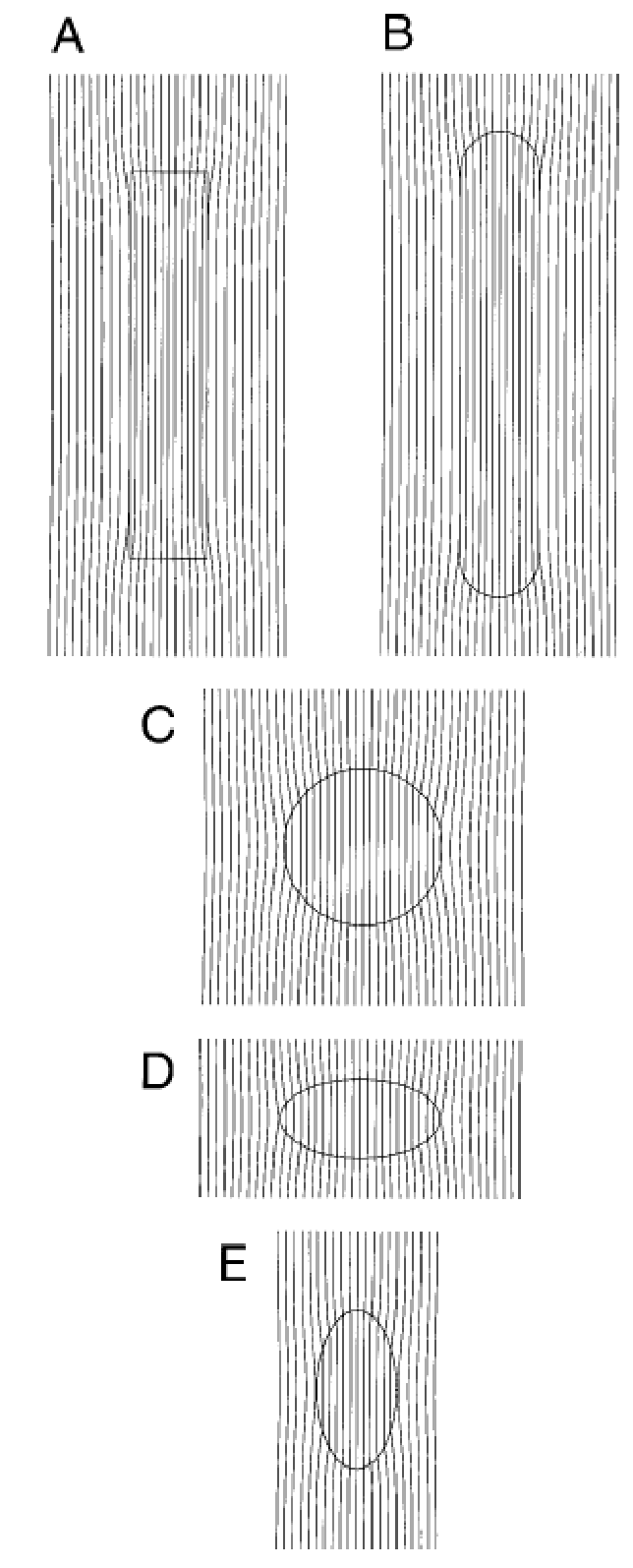}
\caption{Magnetic field lines in and around five different
geometric bodies that have axial symmetry in the direction
of a previously uniform imposed magnetic field. The fields
were calculated using the theory encompassed in Eqs. [43]--[48] 
of the accompanying article (22). The internal magnetic
susceptibility used in the calculations was set to 0.5 (to
make the field distortion visible) and the external susceptibility
was 0. Of course, only the difference in susceptibility
is significant in these calculations. For cylinders (A) and
(B), the length of the straight sides was 10 times the radius
of cross-section. For the oblate spheroid (D), and prolate
spheroid (E), the semimajor axis was twice the semiminor
axis, and the semimajor axes were the same as the radius of
the sphere, (C).}
\end{figure}

Magnetic field lines in and around five different
geometric bodies that have axial symmetry in the direction
of a previously uniform imposed magnetic field. The fields
were calculated using the theory encompassed in Eqs. [43]--[48] 
of the accompanying article (22). The internal magnetic
susceptibility used in the calculations was set to 0.5 (to
make the field distortion visible) and the external susceptibility
was 0. Of course, only the difference in susceptibility
is significant in these calculations. For cylinders (A) and
(B), the length of the straight sides was 10 times the radius
of cross-section. For the oblate spheroid (D), and prolate
spheroid (E), the semimajor axis was twice the semiminor
axis, and the semimajor axes were the same as the radius of
the sphere, (C).

\subsection*{Laplace Equation: Solution}
Mathematical expressions for the macroscopic field
inside and around any body are derived by several
possible means but most directly by solving the
Laplace equation for the magnetic potential (e.g., 23).
It is known from one of the four Maxwell equations
(Ampere's law with ${\bf J}=0$; 24) that the curl of the
magnetic field in a magnetostatic situation is zero, so
$\nabla\times{\bf H}=0$, hence, ${\bf B}$ is described by the gradient of
a scalar potential such that ${\bf H}=-\nabla\phi$. Then, from Eq. [2] ${\bf B}$ can also be found from the scalar potential viz., ${\bf B}=-\mu\nabla\phi$. 
Another of Maxwell's equations (absence of free magnetic monopoles) states that $\nabla\cdot{\bf B}=0$, so
\begin{subequations}
\begin{eqnarray}
\nabla\cdot(\mu\nabla\phi)=0
\end{eqnarray}
In uniform materials, for which $\mu$
is a constant value,
this equation reduces to the Laplace equation:
\begin{eqnarray}
\nabla\cdot\nabla\phi=0
\end{eqnarray}
\end{subequations}
In turn, in Cartesian coordinates this equation is written as
\begin{eqnarray}
\frac{\partial^2\phi}{\partial x^2}+
\frac{\partial^2\phi}{\partial y^2}+
\frac{\partial^2\phi}{\partial z^2}=0
\end{eqnarray}
The solution of the Laplace equation for a given
boundary/body depends on the specification of the
behaviour of $\phi$ at infinity and at the surface of the
body (23). The latter depends on the magnetic susceptibilities
inside and outside the body. For bodies of
various shapes the solution of the Laplace equation
entails finding a coordinate system for which the body
surface is a coordinate surface, in which case Eq. [8]
takes on a much more complicated form. This new
representation of the Laplace equation enables the use
of the mathematical method of separation of variables
to solve it (e.g., 15, 23).

\subsection*{Special Cases}
It is a well-established theoretical prediction and an
experimentally verified fact that if a homogeneous
spherical body is placed in a uniform imposed magnetic
field then the resulting field inside the body is
uniform, even if the magnetic susceptibility of the
material is different from that outside (see references
in 15, 19, 24). This is also the outcome for oblate and
prolate spheroids and even for general ellipsoids.
Thus, it occurs with bodies described as degree-2
surfaces in Cartesian coordinates and is a mathematical
result that can be traced to the fact that the
solution of a second-order differential equation has an
indicial equation of degree 2 (23). Interestingly, a
uniform field also arises in the central spheroid in a
series of confocal spheroids (15). In all cases involving
ellipsoids the direction of the field is not parallel to the
imposed field if the axis/axes of rotational symmetry are
not parallel to the imposed uniform field.

For a long homogeneous cylinder in a uniform
imposed field, the field inside is uniform apart from
inhomogeneities near the two ends [Fig. 2(A)]. Thus,
the cylinder behaves like an elongated prolate spheroid
and the internal field is only parallel to its long
axis if it is parallel to the imposed field.

Analysis yields an expression that provides the
value of the macroscopic field inside the body, but it
does not specify the field at the level of an atomic
nucleus in a molecule in the body. After all, it is the
latter field that determines the Larmor frequency of
the nucleus so it is the one whose value we seek in
order to predict the Larmor frequency of the nucleus
when it is inside the body. The subtleties of calculating
this nuclear field, ${\bf B}_{\mbox{\tiny{nuc}}}$, 
are presented in the accompanying
article. It suffices for this article to simply
declare that the value of ${\bf B}_{\mbox{\tiny{nuc}}}$
in a spherical body, even
when its magnetic susceptibility is different from outside,
is the same as that of the uniform imposed field
outside. On the other hand, for a long cylinder the value
of $\Bnuc$ is different from $\BO$. It is greater or less than $\BO$
depending upon whether the body has a greater or lesser
magnetic susceptibility than outside and on its orientation
with respect to the direction of $\BO$.
Manipulation of these macroscopic situations forms
the basis of an elegantly simple means of measuring the
magnetic susceptibility of a solution, as follows.

\section*{MEASURING MAGNETIC SUSCEPTIBILITY}

\subsection*{Apparatus}
This measurement is most conveniently performed in
a modern NMR spectrometer by using the method of
Frei and Bernstein (25). It employs a glass capillary
that is expanded out at one end to a small sphere to
make a capillary--sphere ({\emph cs}) assembly (Fig. 3). 
\begin{figure}[tb]
\includegraphics[width=\columnwidth]{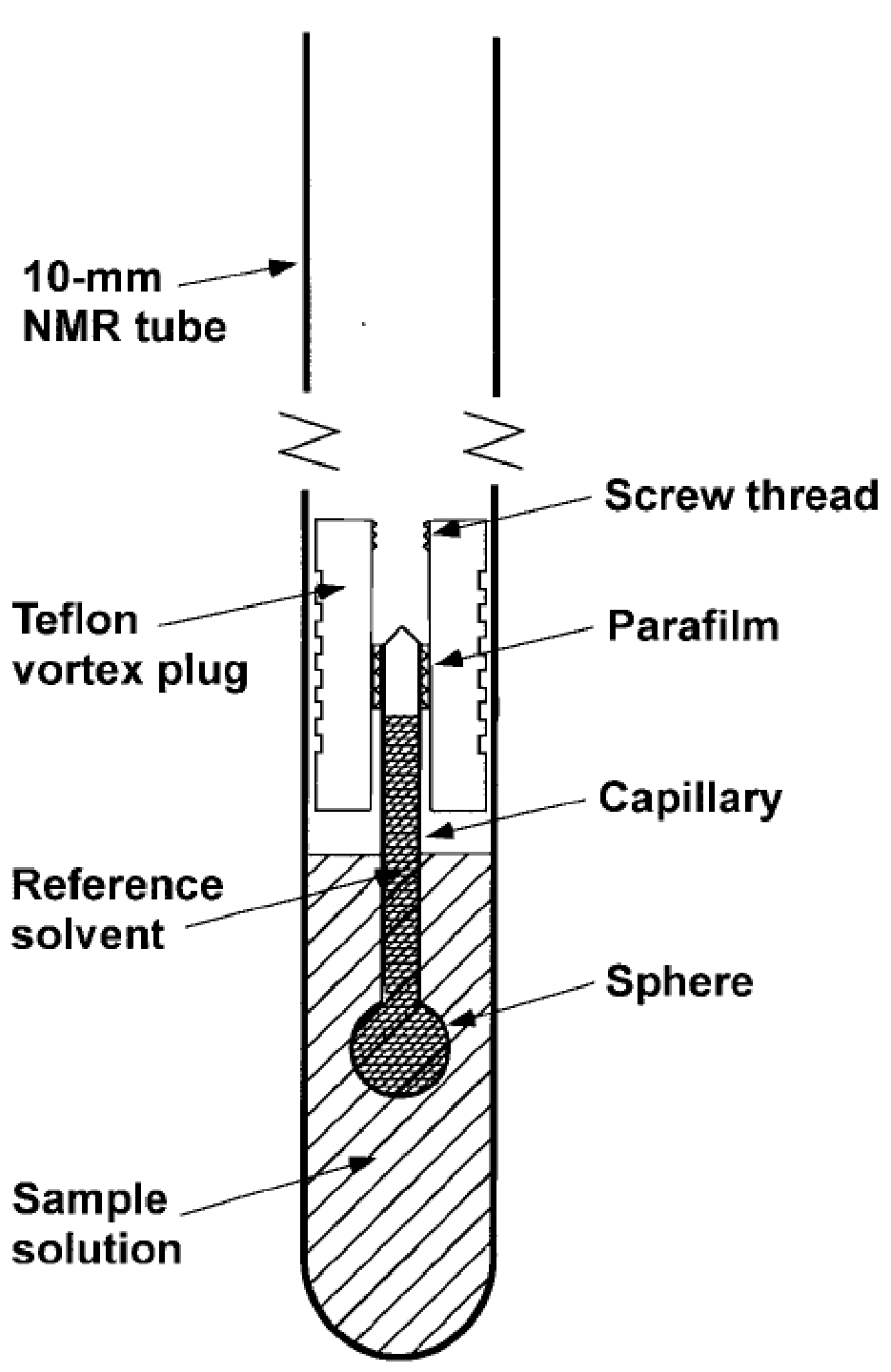}
\caption{Capillary–sphere (cs) assembly for the measurement
of magnetic susceptibility using NMR spectroscopy.
The sample solution whose magnetic susceptibility is to be
determined is placed in the 10-mm NMR tube before insertion
of the {\emph cs} and vortex plug. The cs contains a reference
liquid, such as benzene. The {\emph cs} is positioned in the sample
tube so that signals of similar intensity are obtained from the
contents of both the capillary and the sphere.}
\end{figure}
These devices are commercially available from, e.g., Wilmad
(Buena, NJ; catalog item 529A). The capillary
has an internal diameter of $\sim 1$ mm, an external diameter
of 1.5 mm, and is 40 mm long; the sphere has
an external diameter of 4.1 mm. The reference compound
that is commonly used in the {\emph cs} is benzene
because it has a single $^1$H NMR resonance that is
separated from those of many (biologic) compounds.
Melting the glass at the top of the capillary seals the
{\emph cs}; for nonorganic solvents the tube can be sealed with
Parafilm. The hole of a Teflon vortex plug is drilled
out to an internal diameter of 2.5 mm to accommodate
the capillary, which is held in place with Parafilm
wrapped around its upper end. The flexibility of the
Parafilm enables the sphere to be readily adjusted to
lie coaxially in the sample tube.

We use a conventional 10-mm NMR tube to hold
3 mL of the liquid whose susceptibility is to be
measured. The vortex plug, with the {\emph cs} inserted into it,
is positioned in the NMR tube so that the sphere lies
within the lower limit of the receiver coil, as judged
by using the sample depth gauge that is supplied by
the NMR probe manufacturer. This choice of position
ensures that a signal is readily detected from the
contents of {\emph both} the capillary and the sphere.

A simple 90$^\circ$ pulse-acquire RF-pulse sequence is
used to record a spectrum that, of course, includes
resonances from the benzene in the {\emph cs} and from the
compounds in the sample. Other reference solvents or
solutes in solution, for which the magnetic susceptibilities
are known, can be used and other nuclides
such as $^{13}$C, $^{19}$F, or $^{31}$P, detected.

\subsection*{Analysis: Example}
Figure 4(A) shows a typical $^1$H NMR spectrum from
\begin{figure}[tb]
\includegraphics[width=\columnwidth]{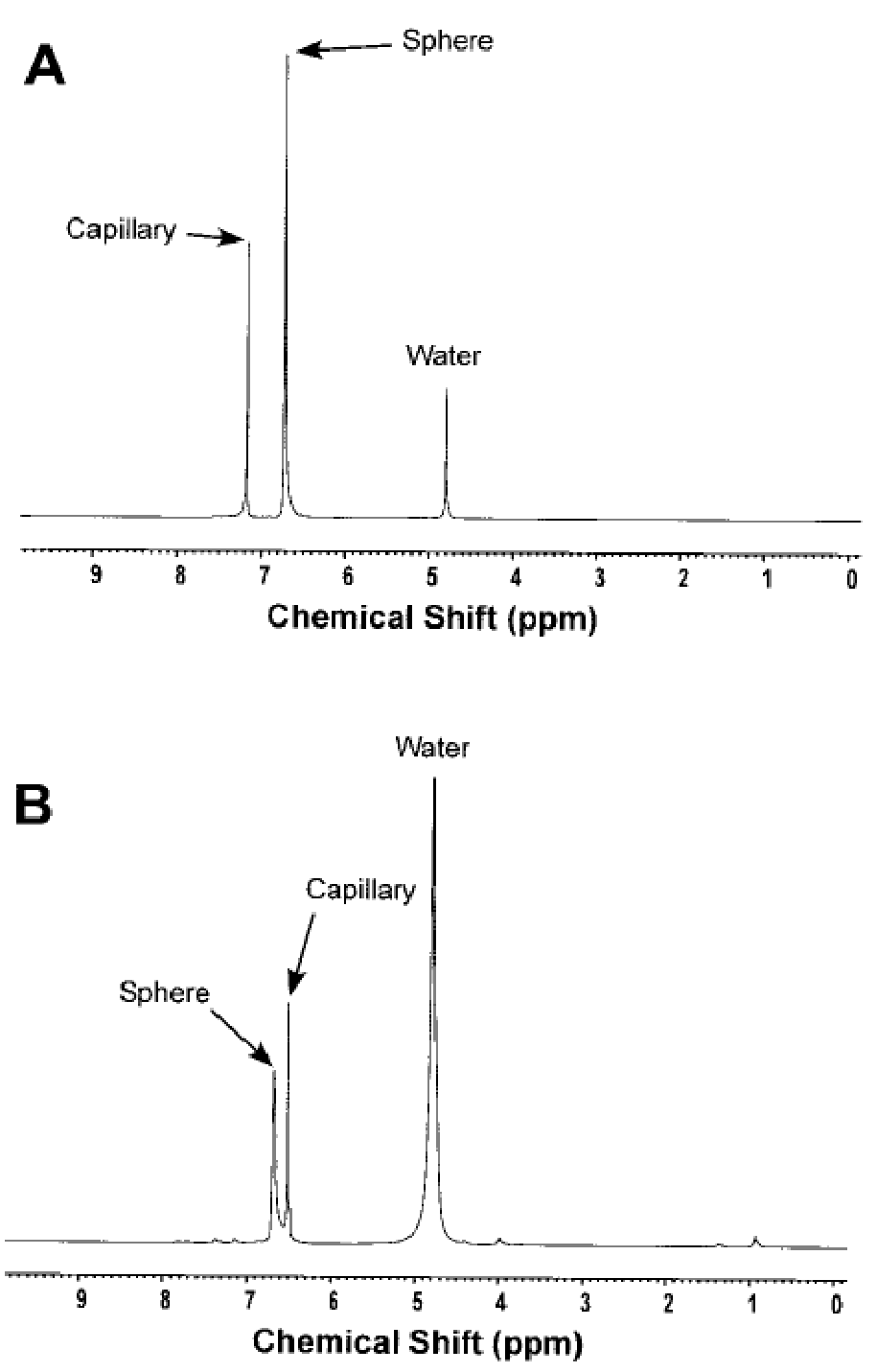}
\caption{$^1$H NMR spectra obtained in a Frei–Bernstein
experiment to measure the magnetic susceptibility of a
liquid sample. The {\emph cs} contained neat benzene and for (A)
the sample was $^2$H$_2$O. For (B) the sample was $^2$H$_2$O to
which had been added iron--dextran (Sigma; 15 mg mL$^{-1}$)
and diethyl phthalate (DEP) (15 mmol L$^{-1}$); the latter gave
rise to the small resonances at approximately 0.9, 1.35, 3.95,
4.4, 7.15, and 7.35 ppm. For these, and all other spectra
presented, the NMR spectrometer was a Bruker DRX 400
with an Oxford Instruments wide-bore vertical magnet, with
the variable temperature unit set to 25$^\circ$C; and, the spectra
were acquired with the simple delay--$\pi/2$-acquire RF-pulse
sequence, with a spectral width of 4 kHz and repetition time
of 0.8 s, and 32 transients were summed for each spectrum.}
\end{figure}
a Frei--Bernstein experiment for which there was benzene
in the cs and $^2$H$_2$O in the sample tube. The signal
labeled ``water" is from a small amount of $^1$HO$^2$H in
the $^2$H$_2$O, while the assignments to the benzene in the
sphere and the capillary were based on the relative
volumes of each in the region of the receiver coil;
these are readily determined by readjusting the position
of the cs with respect to the receiver coil and then
recording another spectrum.

The estimated value of $\chi$ of the sample ($^2$H$_2$O)
depends linearly on the separation between the two
resonances and is given by (25)
\begin{eqnarray}
&&\delta_{\mbox{\tiny{cyl}}}(\mbox{benzene})-
\delta_{\mbox{\tiny{sph}}}(\mbox{benzene})\nonumber\\
&&\,\,\,\,\,\,\,\,\,\,\,\,\,\,\,
=(g_{\mbox{\tiny{cyl}}}-g_{\mbox{\tiny{sph}}})
[\chi(\mbox{benzene})-\chi(^2{\mbox H}_2{\mbox O})]\,\,\,\,\,\,\,\,\,\,\,
\end{eqnarray}
where $\delta$ is the chemical shift (in ppm, measured with
a standard reference compound or, as was done here,
assigning the chemical shift of water to 4.8 ppm), the
$\chi$s denote the magnetic susceptibilities, and the $g$s are
geometric constants that depend on the shape of the
compartment. In the particular case of a cylinder lying
parallel to $\BO$, and a sphere, the factors are -1/3 and
0, respectively (see Eq. [51] in the accompanying
article (22), for which $g=\mathcal{D}_s-1$). The sphere and
capillary are rarely of perfectly ideal shape, so the
term ($g_{\mbox{\tiny{cyl}}}-g_{\mbox{\tiny{sph}}})=G$ is determined as a single
calibration factor using the known susceptibilities of
benzene and $^2$H$_2$O (21). In the present case, therefore,
Fig. 4(A) can be used to estimate $G$.
Thus, 
$\delta_{\mbox{\tiny{cyl}}}(\mbox{benzene})-\delta_{\mbox{\tiny{sph}}}(\mbox{benzene})$ 
was measured to be 0.455 ppm (182 Hz at 400 MHz; see
caption of Fig. 4). The values of 
$\chi({\mbox benzene}) =-6.13\times 10^{-7}$ and 
$\chi(^2\mbox{H}_2\mbox{O})=-7.02\times 10^{-7}$ were
obtained from a table of molar susceptibilities that is
comprehensive but they are given in cgs-emu units
(21). Magnetic susceptibility values of compounds
that are useful in biologic work are given in Table 1.
The values were converted to SI units by using the
multiplicative factor $4\pi\times 10^{-3}$ or, if we continue to
express density of matter in units of g cm$^{-3}$, we use
$4\pi$. Thus,
\begin{eqnarray}
G\amp=\amp [
\delta_{\mbox{\tiny{cyl}}}(\mbox{benzene})-
\delta_{\mbox{\tiny{sph}}}(\mbox{benzene})]\nonumber\\
&&\,\,\,\,\,\div
[\chi(\mbox{benzene})-\chi(^2{\mbox H}_2{\mbox O})]\nonumber\\
&=&[0.455\times 10^{-6}]/(4\pi\times 10^{-7}[-6.11+7.02])\nonumber\\
&=& 0.40
\end{eqnarray}
Hence, for the particular {\emph cs} used for Fig. 4(A), the
value of $G$ was estimated to be 0.40, whereas for an
ideal {\emph cs} the value would have been 0.333. However,
Fig. 4(A) shows only one spectrum from what in
practice was a series of replicated measurements that
were used to estimate a mean value and standard
deviation. In this more extensive series of experiments
with the solvents $^2$H$_2$O, $^1$H$_2$O, methanol, ethanol,
acetone, and CCl$_4$ (see Table 1 for the $\chi$  values
used) the value for $G$ was found to be $0.34\pm0.02$,
whereas for two other $cs$ assemblies the $G$ values were
$0.378\pm0.05$, and $0.374\pm0.04$. This result underscores
the need to calibrate each individual $cs$ assembly
with a number of substances of known $\chi$.
\begin{table}
\caption{Magnetic Susceptibilities of Substances
That Are Useful for Measuring Magnetic
Susceptibilities in Biologic Systems Using NMR
Spectroscopy}
\begin{tabular}{lr}
\hline
Compound Name & 
\begin{tabular}{c}
$\chi(-10^6\times$SI Units:\\
Dimensionless)\footnote{Data were obtained from (26) except for dimethyl sulfoxide,
which was obtained from (10). NB discrepancies exist between the
data for D$_2$O and ethylene glycol in these two information sources.}\\
(20$^\circ$)
\end{tabular}\\
\hline
Acetone & 5.78\\
Benzene & 7.68\\
Carbon tetrachloride & 8.68\\
Dimethyl sulfoxide & 8.55\\
D$_2$O & 8.82\\
Ethanol & 7.23\\
Ethylene glycol & 8.77\\
D-Glucose & (25$^\circ$C) 10.92\\
Glycerol & 9.79\\
H$_2$O & 9.04\\
Mannitol & 11.40\\
Methanol & 6.66\\
Myristic acid & (60$^\circ$C) 8.31\\
Oleic acid & (18$^\circ$C) 8.31\\
Palmitic acid & (62$^\circ$C) 8.31\\
Toluene & 7.76\\
\hline
\end{tabular}
\end{table}

For Fig. 4(B) a paramagnetic reagent, ironÐdextran
containing Fe(III), which is used clinically as an intravenous
iron supplement, was added to the $^2$H$_2$O,
and the same $cs$ as for Fig. 4(A) was inserted into the
sample. In the spectrum, the resonance of the benzene
in the capillary lies to low frequency of the sphere
compared with that in Fig. 4(A) (the other small
resonances that are perceptible in the spectrum are
from diethyl phthalate (DEP), which was also added
to the sample; see below for further discussion of this
point). The separation between the two benzene peaks
is  -0.169 ppm. Hence, by using the value of $G$
determined above, and by rearranging Eq. [10], we
can estimate the $\chi$ of the ironÐdextran solution to be
\begin{subequations}
\begin{eqnarray}
\delta_{\mbox{\tiny{cyl}}}(\mbox{benzene})&-&
\delta_{\mbox{\tiny{sph}}}(\mbox{benzene})/G-\chi(\mbox{benzene})\nonumber\\
&=&-\chi(^2\mbox{H}_2\mbox{O})
\end{eqnarray}
\begin{eqnarray}
\chi(\mbox{Iron}&&\!\!\!\!\!\!\!\!\!\!\mbox{dextran solution})
=-[\delta_{\mbox{\tiny{cyl}}}(\mbox{benzene})\nonumber\\
&&-\delta_{\mbox{\tiny{sph}}}(\mbox{benzene})]/G+\chi(\mbox{benzene})\nonumber\\
&=&-0.169\times10^{-6}/0.34+4\pi(-6.13\times10^{-7})\nonumber\\
&=&4\pi(-5.73\times10^{-7})
\end{eqnarray}
\end{subequations}
The coefficient $4\pi$ has been factored out, so the
bracketed term, which is the value given in cgs-emu
units, can be compared readily with the commonly
used tables (24). The value is clearly negative but less
so than that of neat $^2$H$_2$O.

Diamagnetic samples (e.g., Table 1) have negative
magnetic susceptibilities while paramagnetic ones are
positive, so a mixture of para- and diamagnetic substances
will have a net value that is a weighted sum of
each of the contributions. In the sample used for Fig.
4(B) ironÐdextran was dissolved in $^2$H$_2$O. As we have
seen, $^2$H$_2$O alone is diamagnetic but is ironÐdextran
paramagnetic or simply less diamagnetic? In fact, the
sample has both dia- and paramagnetic constituents so
a key question is, how do we measure the magnetic
susceptibility of each constituent of a mixture, and in
turn how do we use the values to predict the net value
for a mixture? We now address this task.

\section*{WIEDEMANNÕS ADDITIVITY LAW}

\subsection*{Different $\chi$-Types}
This law (19) is a consequence of the superposition
principle for electrostatic and magnetic fields that
specifies that the field at a given point in space is the
linear vector sum of contributions from all sources.
WiedemannÕs law states that the overall magnetic
susceptibility of a mixture is the weighted sum of the
magnetic susceptibilities of the constituents, weighted
according to their relative volumes of occupation of
the mixture, which is usually a solution. Thus, in
mathematical form it is written as
\begin{eqnarray}
\chi(\mbox{mixture})=\sum_{i=1}^N V_i\chi_i\Bigg{/}\sum_{i=1}^N V_i
\end{eqnarray}
where $V_i$ denotes the volume of the mixture occupied
by substance $i$, whose magnetic susceptibility is $\chi_i$.
Therefore, it is evident why the susceptibility, which
is dimensionless, is nevertheless referred to as the
volume susceptibility and often written with a subscript
$V$, namely, $\chi_{V,i}$.

On the other hand, it is sometimes useful to express
the susceptibilities in terms of masses; hence in a
mixture, the weighting factors are the corresponding
masses:
\begin{eqnarray}
\chi(\mbox{mixture})=\sum_{i=1}^N m_i\chi_{\mbox{\tiny{mass}},i}\Bigg{/}\sum_{i=1}^N m_i
\end{eqnarray}
where $\chi_{\mbox{mass,i}}$ is the so-called mass susceptibility and
the $m_i$ are the relative masses of the components of the
mixture. Because $m_i= \rho_i V_i$, where $\rho_i$ is the density of
the $i$th component, the relationship between mass and
volume susceptibility is
\begin{eqnarray}
\chi_{\mbox{\tiny{mass}},i}=\chi_{\mbox{\tiny{vol}},i}\bar{\nu}_i
\end{eqnarray}
where for convenience we use the reciprocal of the
density, $\bar{\nu}_i$, known as the partial specific volume.

Yet another way of expressing the overall magnetic
susceptibility of a mixture is to use the number
of moles, $n_i$, of each substance in the mixture; this
uses molar magnetic susceptibilities as follows:
\begin{eqnarray}
\chi(\mbox{mixture})=\sum_{i=1}^N n_i\chi_{\mbox{\tiny{mol}},i}\Bigg{/}\sum_{i=1}^N n_i
\end{eqnarray}     
Thus, the molar susceptibility is related to the volume
susceptibility by
\begin{eqnarray}
\chi_{\mbox{\tiny{mol}},i}=m w_i \bar{\nu}_i\chi_{\mbox{\tiny{vol}},i}
\end{eqnarray}
As noted above, tables of magnetic susceptibilities are
often given as molar susceptibilities (e.g., 26) but
when studying solutions it is simplest to use volume
susceptibilities, as given in Table 1. As well as being
cautious with the factor of $4\pi$, care must be exercised
in correcting for changes in solute density with temperature.

\subsection*{Complications with WiedemannÕs Law}
There are potential traps in calculating the net magnetic
susceptibility of a mixture from the known susceptibilities
of all the constituents and their relative
volumes or masses. The most obvious problems arise
if there are chemical reactions between the constituents;
any changes in chemical properties can clearly
change magnetic ones. Another effector of susceptibility
is a change in the conformation of a macromolecule,
such as occurs in hemoglobin when ligands
such as 2,3-bisphosphoglycerate bind to it; and, a
further effect is brought about by the binding of
oxygen to heme that alters the spin state of the prosthetic
Fe atom (27, 28).

\section*{RED BLOOD CELLS}

\subsection*{Measurement of $\chi$}  
Figure 5(A) shows the $^1$H NMR spectrum from a
FreiÐBernstein experiment in which the sample was a
suspension of RBCs of hematocrit (Ht) 77\%, and the
central peak is from the residual $^1$H$_2$O in the cells that
had been centrifugally washed in $^2$H$_2$OÐsaline. In
contrast, Fig. 5(B) shows the spectrum obtained with
the lower Ht of 54\%. 
The larger separation (200 Hz
compared with 183 Hz) between the benzene resonances
from the cs for the suspension of lower Ht
indicates that the RBCs were less diamagnetic than
the saline bathing medium.
\begin{figure*}[tb]
\includegraphics[width=150mm]{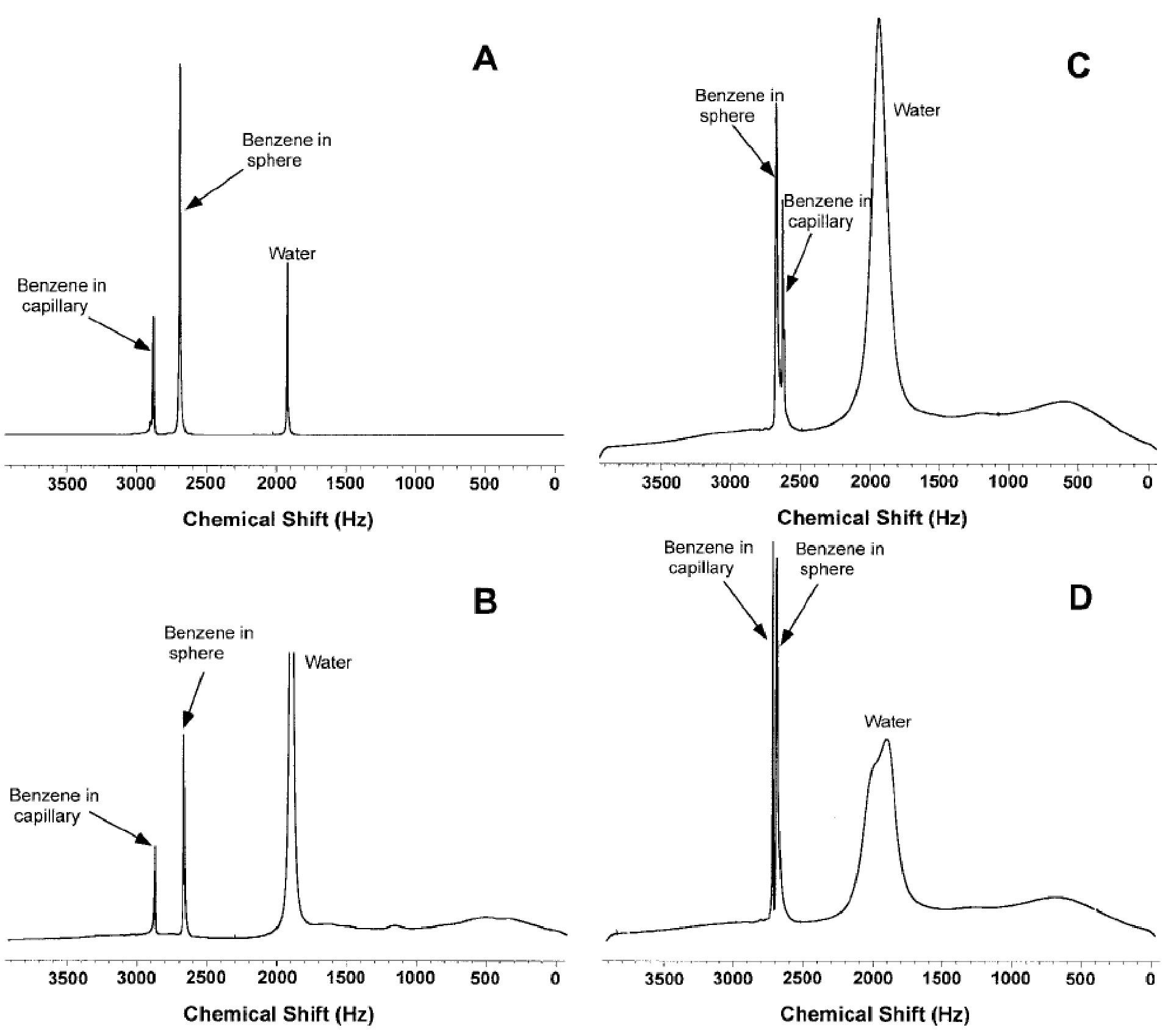}
\caption{$^1$H NMR spectra from a FreiÐBernstein experiment conducted on suspensions of RBCs.
The reference solution was benzene. The RBCs were obtained freshly by venipuncture of the median
cubital vein from a healthy volunteer (P.W.K.). The RBCs were washed three times in isotonic
saline (154 mmol L$^{-1}$ NaCl) according to a standard protocol (13). 
(A) Oxygenated RBCs in a suspension of Ht = 77\%. 
(B) Oxygenated RBCs in a suspension of Ht = 54\%. 
(C) RBCs prepared
first as for (A) but then NaNO2 was added in a 1.5:1 stoichiometric ratio, with the cells turning
brown over a period of  2 min. The Ht value was 77\%. 
(D) RBCs prepared as for (C) but with Ht = 54\%.}
\end{figure*}

Further, Fig. 5(C) shows the outcome from RBCs
of Ht   77\%, 
in which the Fe in the hemoglobin was
oxidized by adding NaNO$_2$ to the suspension (29). In
the presence of molecular oxygen the nitrite abstracts
an electron from Fe(II) in the heme of hemoglobin to
generate nitrate and an Fe(III)Ðhemoglobin complex
that is known to be paramagnetic. The resonance from
the benzene in the cylinder is now to low frequency of
that from the sphere. However, for Ht = 54\% 
[Fig. 5(D)] the magnitude of the paramagnetic effect of the
Fe(III) in the hemoglobin is weaker so that the resonance
from the capillary remains (slightly) to high
frequency of the peak from the sphere.

An important conclusion that can be drawn from
these data is that the magnetic susceptibility of an
RBC suspension of hematocrit in the physiological
range of 35Ð45\%, 
which contains all its hemoglobin
in the (oxidized) met state, is still diamagnetic. The
net magnetic susceptibility of a suspension of RBCs is
only positive, namely, paramagnetic, if all the hemoglobin
is converted to the met form and the Ht is well
above the physiological value of  $\sim$40\%.

The net magnetic susceptibility of the RBC suspension
is the volume-weighted sum of the volume
susceptibility of water, ions, membrane constituents,
and the hemoglobin molecules (which make up 95\%
of the cellÕs proteins). In turn, the magnetic susceptibility
of hemoglobin is the weighted sum of the diamagnetic
susceptibility of the globin, the diamagnetic
susceptibility of the heme moieties, and the paramagnetic
susceptibility of Fe(III), as has been reported by
Cerdonio et al. (27, 28).

Another obvious finding from Figs. 5(C) and (D) is
the broadness of the water signal in the presence of
paramagnetic hemoglobin. This is understood to be
due to two effects:
\begin{enumerate}
\item{The large difference in the magnetic susceptibility
between the inside and outside of the cells
creates large spatial magnetic field variations in
and around the cells. This gives rise to a distribution
of Larmor frequencies.}
\item{Rapid exchange between free water and that
associated with hemoglobin ensures that the
high-energy nuclear magnetic state is relaxed
rapidly by the paramagnetic Fe(III) in the hemoglobin.
Further, water is in rapid exchange
across the cell membranes, in a process mediated
primarily by aquaporins, so protons in the
water molecules outside the cells also have their
relaxation rates enhanced.}
\end{enumerate}

\section*{EMULSIONS AND SOLUTIONS}

\subsection*{Context}
We recently found that some amphipathic molecules,
in particular various phthalate esters, when added to
suspensions of RBCs give rise to two sets of $^1$H NMR
resonances. The study was part of an investigation of
the drug detoxification characteristics of human RBCs
and the phthalates were used as model xenobiotic
compounds (30). Diethyl phthalate in a suspension of
RBCs not only gave two sets of resonances from each
proton in the molecule but over time the relative
intensities of one of the sets declined (30, 31). The
separate sets of resonances were attributed to intraand
extracellular populations of the compound, while
the change in relative intensity with time was originally
attributed to metabolism. Only when the $^1$H
NMR spectrum of the stock aqueous sample was
obtained was it appreciated that the two sets of peaks
were from DEP in free solution, and the remainder
was in (presumably spherical) microdroplets. The explanation
for the decline in one set of peaks was the
binding of the free compounds to cellular proteins,
probably hemoglobin, and the progressive coalescence
of the microdroplets in a process of separation
of the phthalate phase from the aqueous one.

The analysis of these spectra was illuminated by an
understanding of the effects on the proton resonance
frequencies of having neat DEP in microdroplets. The
significantly different magnetic susceptibility of the
DEP in aqueous solution and in the microdroplets was
at least part of the basis of the different resonance
frequencies. The various features of the spectra and a
more general understanding of the effects of differences
in magnetic susceptibility in heterogeneous systems
that arise in the DEP system are described next.

\subsection*{Neat DEP}
Figure 6(A) shows the structure of DEP and its $^1$H
NMR spectrum. The relatively low resolution has a
positive pedagogic outcome: It enables us to focus
attention on the main resonances and not their fine
structure or splitting patterns. Assignment of the
methyl and methylene resonances is straightforward,
with the methylene resonance at the higher frequency.
\begin{figure}[tb]
\includegraphics[width=\columnwidth]{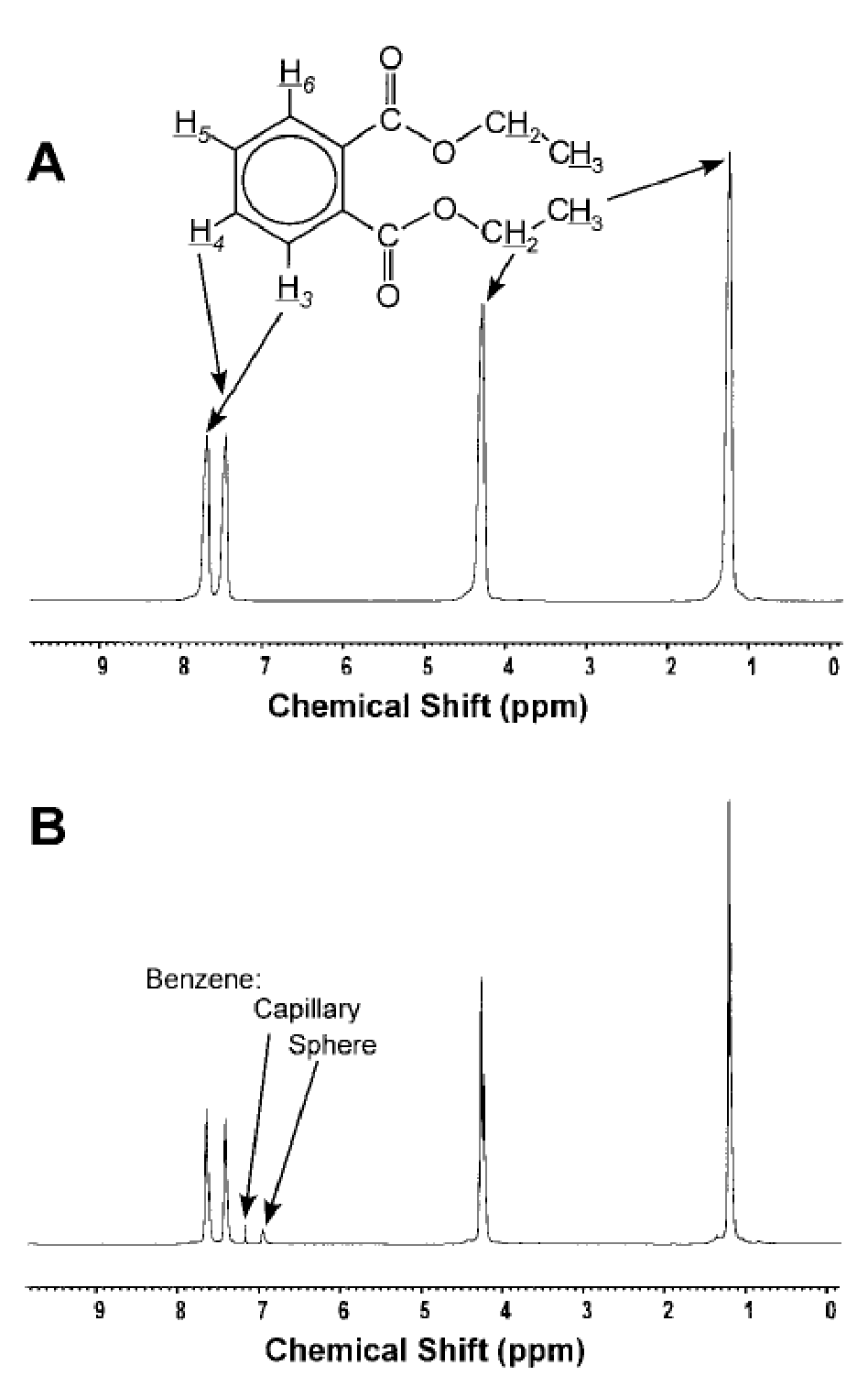}
\caption{(A) Low-resolution $^1$H NMR spectrum of neat
DEP in a 10-mm glass NMR tube and (B) spectrum from a
FreiÐBernstein experiment with DEP in a 10-mm NMR
sample tube and neat benzene in the \emph{cs}.}
\end{figure}
However, because proton chemical shifts are not
readily predicted for aromatic rings with two \emph{ortho}
substituents (32), and the solvent system we describe
here might be considered somewhat unusual, we confirmed
the assignments shown in Fig. 6(A) by HSQC
and HMBC spectra. Note that in Fig. 6(A) the position
numbers of the atoms are given in italics and do not
indicate spin systems. The spin systems are of the
type AA'BB'  for each set of aromatic protons. The
HMBC spectrum yielded a strong correlation due to
the 3-bond coupling between the protons that are
\emph{ortho} to each carboxyl group and the carboxyl carbon
atoms; no correlation was observed for the relatively
small coupling between the \emph{meta} protons and carboxyl
carbon atom.

The relative areas of the peaks corresponding to
the protons in DEP should be in the ratio 3:2:2:2 for
the respective signals from low to high frequency.
However, this is clearly not the case. The ratio is
approximately correct for the aliphatic protons, but
the aromatic protons, which have longer $T_1$ values,
did not fully relax between transients because the
recycle time of spectral acquisition (compared with
$T_1$) was insufficient.

\subsection*{$\chi$ of DEP}
Figure 6(B) shows the $^1$H NMR spectrum from a
FreiÐBernstein experiment with neat DEP in the sample
tube giving slightly better spectral resolution to
that shown in Fig. 6(A); we used the same benzene-containing
\emph{cs} assembly described above. The signal
from the benzene in the capillary is at higher frequency
than from that in the sphere. This indicates
that neat DEP is diamagnetic. In fact, from the difference
in chemical shift of 0.21 ppm (84.1 Hz) we
can use Eq. [11] to obtain an estimate of the volume
magnetic susceptibility of  $-4\pi\times6.62\times10^{-7}$. The
negative sign indicates that DEP is diamagnetic.

\subsection*{$^2$H$_2$O as Sample and DEP in \emph{cs}}
Figure 7 shows the $^1$H NMR spectrum from a FreiÐ
Bernstein experiment in which the sample was $^2$H$_2$O
and DEP was in the \emph{cs} instead of benzene. Deuterated
water was used to reduce the intensity of the $^1$H$_2$O
peak. 
\begin{figure}[tb]
\includegraphics[width=\columnwidth]{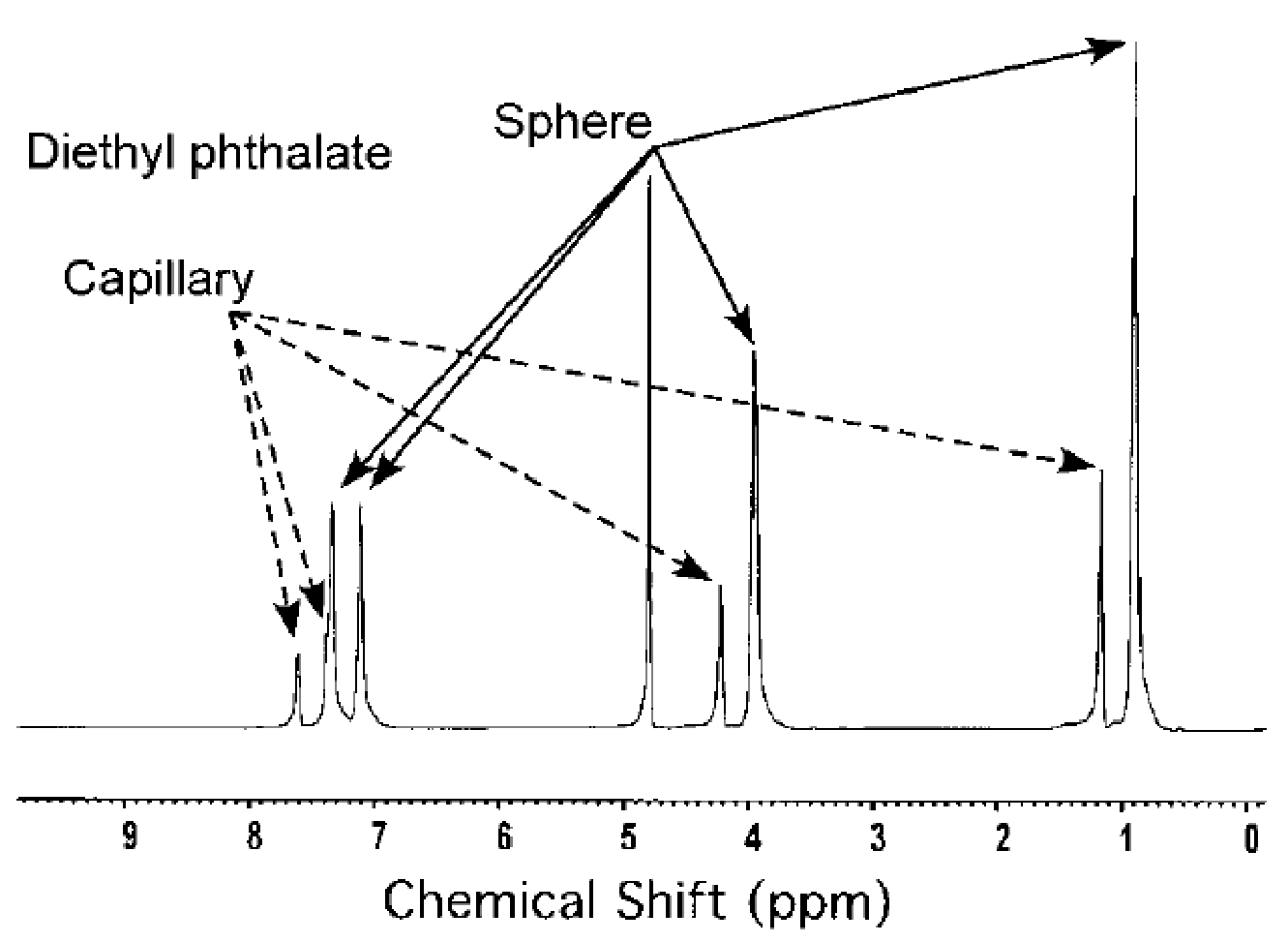}
\caption{$^1$H NMR spectrum of neat DEP in a glass cs
assembly (see Fig. 3) with $^2$H$_2$O in the NMR tube. The
resonances from DEP in the glass sphere of the \emph{cs} are
indicated by solid arrows, whereas resonances from DEP in
the cylindrical capillary are indicated by the broken arrows.}
\end{figure}
Thus, the spectrum shows a singlet from the
trace of water ($^2$HO$^1$H) in the $^2$H$_2$O sample, with a
chemical shift of  4.8 ppm. For the DEP in the \emph{cs}, the
methyl and methylene protons give clearly resolved
peaks from the capillary and sphere. However, the
peak at $\sim$7.35 ppm from H$_3$/H$_6$ on the benzene ring
of DEP in the capillary fortuitously overlaps with the
peak from the H$_4$/H$_5$ of the DEP in the sphere.

Recall that the only difference between the DEP
that gives the two separate sets of peaks in Fig. 7
compared with Fig. 6 is the fact that part of the sample
for the spectrum in Fig. 7 was located in a glass
capillary and the other was in a glass sphere, while
outside there was a substance of different magnetic
susceptibility, $^2$H$_2$O. The peak separation in Fig. 7 of
0.27 ppm (107 Hz) is substantially greater than, for
example, the chemical shift change of the  -protons
of an amino acid such as glycine in aqueous solution
with a pH change from 5 to 8.

An appreciation of the extent of this magnetic
susceptibility-induced shift is important for the next
stage in the interpretation of the spectra of DEP in an
aqueous environment.

\subsection*{DEP in $^2$H$_2$O in Sample with No \emph{cs}}
The spectrum in Fig. 8(A) was obtained with neat
DEP added to $^2$H$_2$O to give a concentration, averaged
over the sample, of 10 mmol L$^{-1}$; no \emph{cs} was used.
When preparing the sample, the added DEP did not
completely dissolve in the $^2$H$_2$O and phase separation
was obvious, but vigorous shaking produced a slightly
opaque mixture that appeared to be stable, at least to
the eye, for several hours.
\begin{figure}[tb]
\includegraphics[width=\columnwidth]{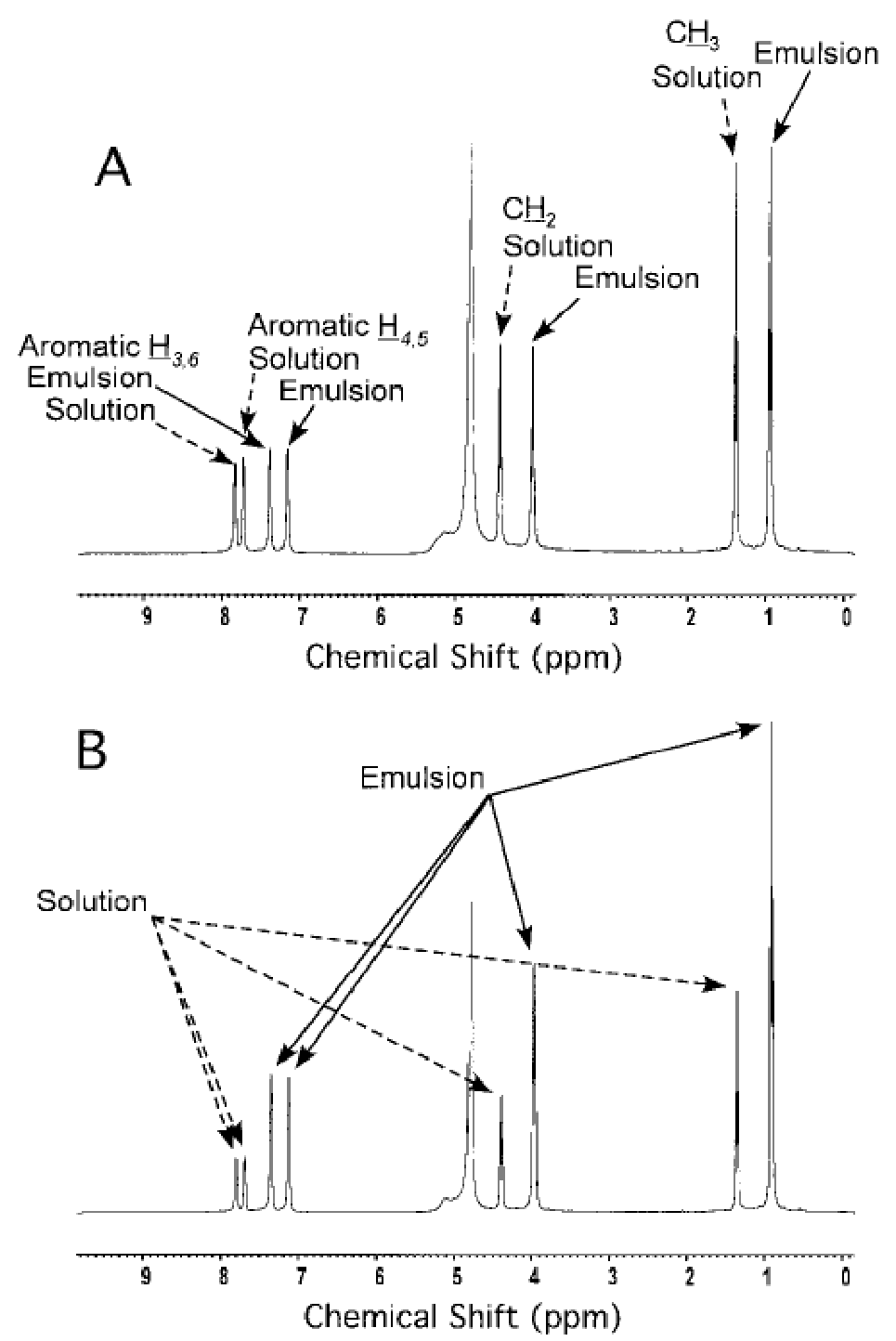}
\caption{$^1$H NMR spectrum of an emulsion of DEP in
$^2$H$_2$O. For resonance assignments, the basis of which is
discussed in the text, see Fig. 6(A). The solid arrows indicate
resonances from DEP in the emulsion phase. The
broken arrows indicate resonances from DEP in aqueous
solution. (A) Overall concentration of DEP was 10 mmol
L$^{-1}$. (B) Overall concentration of DEP was 15 mmol L$^{-1}$.}
\end{figure}

The subsequent $^1$H NMR spectrum contained
twice as many peaks as that of neat DEP, and each
pair of peaks was of almost the same amplitude and
area in the two phases. In fact it was reminiscent of
the spectrum in Fig. 7, for which the DEP was in the
\emph{cs} and $^2$H$_2$O was in the sample tube. As noted in the
Introduction, it was hypothesized that the two sets of
peaks were from two different phases of DEP, one as
neat DEP in the emulsion and one from DEP dissolved
in $^2$H$_2$O. But, which set of peaks is assigned to
the microspheres of the emulsion?

The answer lies in the elegant result (outlined
above) that for an isolated sphere in a uniform imposed
magnetic field $\BO$ the macroscopic field inside
is always uniform and has a value that depends on the
difference in magnetic susceptibility across the
boundary, no matter what the radius of the sphere
with all other things being equal (e.g., solvent effects,
see below; 23, 24). On the other hand, the magnitude
of the magnetic field at a resonant nucleus is the same
as that of a similar nucleus outside the sphere. This
remarkable result is the basis of why a cs assembly is
used in routine practice for external reference compounds
when accurately determining chemical shifts
in NMR spectra. The physical explanation of this
result is the subject of the accompanying article.

The two peaks centered at 7.15 and 7.39 ppm are
separated by 0.24 ppm, which is exactly the same as
the separation between these two peaks from the DEP
in the glass sphere shown in Fig. 7. This fact alone
suggests that the set of four peaks to lower frequency
in Fig. 8(A) are from DEP in the emulsion microspheres.
However, an experiment that yields the assignment
when its spectrum is compared with Fig.
8(A) is shown in Fig. 8(B). The DEP concentration
was increased to 15 mmol L$^{-1}$, and this was accompanied
by an increase in the cloudiness of the mixture.
The resulting $^1$H NMR spectrum showed a doubling
of the amplitudes and areas of the lower-frequency
peaks, thus enabling their assignment to DEP in the
emulsion microspheres. The higher-frequency peaks
are therefore assigned to DEP in aqueous solution,
representing a solubility of $\sim$5 mmol L$^{-1}$ in water, at
25$^\circ$C. In other words, because the limit of solubility of
the DEP was reached at $\sim$5 mmol L$^{-1}$ the more
intense peaks in the spectrum must have been from
the emulsion phase.

A remaining important observation is possible
from Figs. 8(A) and 8(B). The separation between the
resonances of the methyl and methylene protons in the
emulsion phase and the aqueous solution was 0.443
ppm (177.1 Hz) and 0.426 ppm (170.2 Hz), respectively;
those between the corresponding pairs of aromatic
resonances were 0.561 ppm (224.5 Hz) and
0.445 ppm (178.1 Hz), respectively, for the H4/H5 and
H3/H6 pairs. Another way of emphasising the difference
between the solution and emulsion spectra is to
note that the separation between the H$_4$/H$_5$ and H$_3$/H$_6$
resonances is 0.108 ppm (43.1 Hz) in the solution and
0.224 ppm (89.5 Hz) in the emulsion phase.

An explanation for this differential in shifts, between
the emulsion and solution phases, is that H$_3$/H$_6$
are more exposed to the diamagnetic anisotropy of the
carbonyl oxygen of the ester group than are H$_4$/H$_5$. In
the aqueous phase this anisotropy is diminished by the
binding of water. This influence is not readily apparent
for the methyl and methylene protons because the
flexibility of the alkyl chain allows each group of
protons similar proximity to the carbonyl group.

It is clear that solvent effects of this nature can
work either in opposition or additively to magnetic
susceptibility effects to determine the final chemical
shift. This outcome is well illustrated in the next
example.

\subsection*{15 mmol L$^{-1}$ DEP in $^2$H$_2$O and
Neat DEP in \emph{cs}}
The $^1$H NMR spectrum from the sample arrangement
given in the title of this section (Fig. 9) illustrates the
predicted superposition of the resonances of the DEP
in the sphere of the cs and of DEP in the emulsion
phase; the superimposed peaks at 0.87, 3.9, 7.1, and
7.4 ppm are due to DEP in spherical compartments.
\begin{figure}[tb]
\includegraphics[width=\columnwidth]{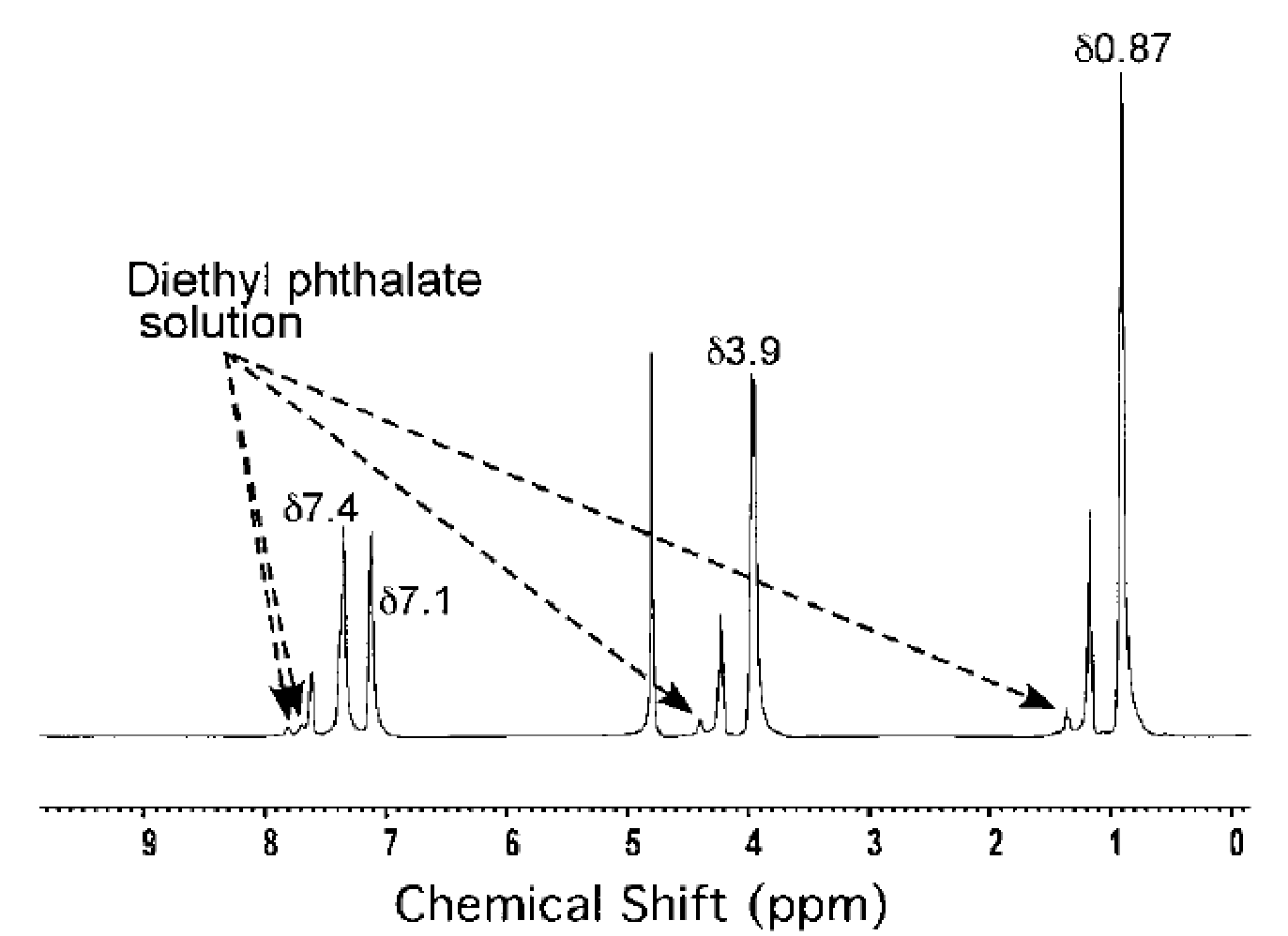}
\caption{$^1$H NMR spectrum of 15 mmol L$^{-1}$ DEP in
$^2$H$_2$O in a 10-mm NMR tube with neat DEP in the \emph{cs}
assembly. The resonances from the DEP in the aqueous
phase, i.e., in solution, are indicated by the broken arrows,
while the assignments of the other resonances are described
in the text.}
\end{figure}

Peaks of about one-third the intensity of those due
to DEP in spherical compartments are found at 1.15,
4.2, 7.4 (unresolved from the highest-frequency
sphere-peak), and 7.65 ppm. These arise from DEP in
the capillary of the cs. Finally, the small peaks identified
by arrows in Fig. 9 are those from DEP dissolved
in the $^2$H$_2$O.

The assignments for DEP in the sphere (and coincident
resonances for DEP in the emulsion phase) and
cylinder of the \emph{cs} were determined by monitoring
signal intensity while moving the \emph{cs}, as described
above for the benzeneÐwater system [Fig. 4(A)]. The
resonances whose frequency is insensitive to the position
of the \emph{cs} may therefore be attributed to DEP in
solution in the (external) NMR tube.

\section*{OTHER PHENOMENA THAT CAUSE
SHIFT EFFECTS}

\subsection*{H-Bonding}
Specific shift effects, other than magnetic susceptibility
differences in a sample, occur with some phosphoryl
and various $^{13}$C and $^{19}$F resonances in RBC suspensions.
These latter effects we have denoted ``split
peak phenomena" (33, 34); they arise from the different
average extent of H-bonding of water inside
and outside the cells to the phosphoryl oxygen, the F
atom, or the oxygen near the reporter $^{13}$C nucleus,
respectively.

\subsection*{Shift Reagents}
Another valuable experimental means of bringing
about a transmembrane NMR chemical shift difference
exists for alkaliÐmetal cations; paramagnetic lanthanide
shift reagents are the main group. More recently
the chemical shifts of inorganic anions have
been altered by cobalt complexes of glycine and triglycine
(6, 35, 36). Thus, the chemical shift of a
solute in a cellular system can be affected by BMS,
solvent, and shift-reagent-induced effects that may be
additive or negate each other (e.g., 32).

\section*{CONCLUSIONS}

The FreiÐBernstein experiment (Figs. 3 and 4) provides
a graphical demonstration of the effect of the
shape of a macroscopic container on the chemical
shift of a nuclear population, when there is a difference
in magnetic susceptibility across the boundary of
the container. The difference in magnetic susceptibility
between that of water and an organic liquid such as
DEP induces a larger shift than those due to the
phenomena mentioned above.

The discovery that differences in magnetic susceptibility
underlie the duplicated resonances of DEP in
dilute aqueous solutions provides insights into otherwise
perplexing data from DEP in RBC suspensions
(30, 31). The BMS effect also accounts for the separate
resonances from lipid in spherical droplets inside
skeletal muscle and outside in elongated, more cylinder-
like, adipocytes (2Ð4, 8). In other words, the
situation with DEP in a capillary (cylinder) and sphere
assembly is an analog of muscle tissue in which there
are lipid droplets inside myocytes and larger, more
elongated, lipid bodies in adipocytes between the fibers.
The lipid is relatively more diamagnetic than the
surrounding medium, and the arm or leg of a patient
will be parallel to $\BO$ of a horizontal magnet in an
MRI/MRS scanner, so the lipid cylinders lie in the
direction of $\BO$; this is the same as the orientation of
the capillary used in our FreiÐBernstein experiment.
Thus, it is possible to predict, from what we presented
above, that the signal from the extramyocellular lipid
will be to high frequency of that from the intramyocellular
lipid; this is confirmed by in vivo MRS (3).

In addition, we emphasized the importance of using
a spherical bulb of the kind shown in Fig. 3 to
contain the reference compound when an external
chemical shift reference is employed in NMR spectroscopy.

While an empirical appreciation of the extent of
BMS effects that might arise in studies of cellular
systems can be obtained from experimental data like
those in Figs 7Ð9 (e.g., 28), a quantitative description,
or prediction of the actual value, is much more complex.
The actual explanation for why the BMS shift
effect does not impinge upon nuclei if they are inside
a spherical bulb entails a subtle argument; this is the
subject of the accompanying article (22).

\section*{ACKNOWLEDGMENTS}

The work was funded by a Project Grant from the
Australian Research Council to P.W.K. Bill Lowe is
thanked for expert technical help and David Philp and
David Regan and Drs. Tom Eykyn and Konstantin
Momot are thanked for valuable discussions on the
article.

\section*{REFERENCES}
\begin{enumerate}
\item{VanderHart DL. Magnetic susceptibility \& high resolution
NMR of liquids \& solids. In: Grant D, Harris R,
eds. Encyclopaedia of NMR. New York: John Wiley \&
Sons. p 2938Ð2946.}
\item{Schick F, Eisman B, Jung WJ, Bongers H, Bunse M,
Lutz O. Comparison of localized proton NMR signals
of skeletal muscle and fat tissue in vivo: two lipid
compartments in muscle tissue. Magn Reson Med
1993; 29:158Ð167.}
\item{Boesch C, Slotboom J, Hoppeler H, Kreis. In vivo
determination of intra-myocellular lipids in human
muscle by means of localized 1H-MR-spectroscopy.
Magn Reson Med 1997; 37:484Ð493.}
\item{Hakumaki JM, Kauppinen RA. $^1$H NMR visible lipids
in the life and death of cells. TIBS 2000; 25:357Ð361.}
\item{Majumdar S, Gore JC. Studies of diffusion in random
fields produced by variations in susceptibility. J Magn
Reson 1988; 78:41Ð55.}
\item{Springer CS. Physicochemical principles influencing
magnetopharmaceuticals. In: Gillies RJ, ed. NMR in
Physiology and Biomedicine. San Diego: Academic
Press; 1994. p. 75Ð99.}
\item{Li L. Magnetic susceptibility quantification for arbitrarily
shaped objects in inhomogeneous fields. Magn
Reson Med 2001; 46:907Ð916.}
\item{Steidle G, Machann J, Claussen CD, Schick F. Separation
of intra- and extramyocellular lipid signals in proton
MR spectra by determination of their magnetic field
distribution. J Magn Reson 2002; 154:228Ð235.}
\item{Homer J, Aldaffaee HK. Volume magnetic-susceptibilities
from nuclear magnetic-resonance chemical-shift
studies of heterogeneously dispersed liquids. J Chem
Soc Faraday Trans 1985; 81:803Ð809.}
\item{Doty DF, Entzminger G, Yang YA. Magnetism in
high-resolution NMR probe design. I: General methods.
Concepts Magn Reson 1998; 10:133Ð156.}
\item{Brindle KM, Brown, FF, Campbell ID, Grathwohl C,
Kuchel, PW. Application of spin echo nuclear magnetic
resonance to whole cell systems: membrane transport.
Biochem J 1979; 180:37Ð44.}
\item{Rabenstein DL. $^1$H NMR method for the non-invasive
study of metabolism and other processes involving
small molecules in intact erythrocytes. J Biophys Biochem
Meth 1984; 8:277Ð306.}
\item{Kuchel PW. Biological applications of NMR. In: Field
LD, Sternhell S, eds. Analytical NMR. Chichester,UK:
John Wiley \& Sons;1989. p 157Ð219.}
\item{Endre ZH, Chapman BE, Kuchel PW. Cell volume
dependence of $^1$H spin echo NMR signals in human
erythrocyte suspensions: influence of in-situ field gradients.
Biochim Biophys Acta 1984; 803:137Ð144.}
\item{Kuchel PW, Fackerell, ED. Parametric-equation representation
of biconcave erythrocytes. Bull Math Biol
1999; 61:209Ð220.}
\item{Kuchel PW, Bulliman BT. Perturbation of homogeneous
magnetic fields by isolated cells modelled as
single and confocal spheroids: implications for magnetic
resonance spectroscopy and imaging. NMR
Biomed 1989; 2:151Ð160.}
\item{Glasel JA, Lee KH. On the interpretation of water
nuclear magnetic resonance relaxation times in heterogeneous
systems. J Am Chem Soc 1974; 96:970Ð978.}
\item{Jones AJ, Kuchel PW. Measurement of choline concentration
and transport in human erythrocytes by $^1$H
NMR: comparison of normal blood and from lithiumtreated
psychiatric patients. Clin Chim Acta 1980; 104:
77Ð85.}
\item{Pople JA, Schneider WG, Bernstein HJ. High-Resolution
Nuclear Magnetic Resonance. New York:
McGraw-Hill; 1959.}
\item{Levitt MH. The signs of frequencies and phases in
NMR. J Magn Reson 1997; 126:164Ð182.}
\item{Levitt MH. Signs of frequencies and phases in NMR:
the role of radiofrequency mixing. J Magn Reson 2000;
142:190Ð194.}
\item{Durrant CJ, Hertzberg MP, Kuchel PW. 2003. Magnetic
susceptibility: further insights into macroscopic
and microscopic fields and the sphere of Lorentz. Concepts
Magn Reson Part A 18A:72Ð95.}
\item{Moon P, Spencer DE. Field Theory for Engineers.
Princeton, NJ: Van Nostrand; 1961.}
\item{Jackson JD. Classical Electrodynamics, 3rd Ed. New
York: John Wiley \& Sons; 1998.}
\item{Frei K, Bernstein HJ. Method for determining magnetic
susceptibilities by NMR. J Chem Phys 1962; 37:1891Ð
1892.}
\item{Weast RC. CRC Handbook of Chemistry and Physics.
Boca Raton, FL: CRC Press; 1983.}
\item{Cerdonio M, Morante S, Vitale S. Magnetic susceptibility
of haemoglobins. Meth Enzymol 1981; 76:354Ð
371.}
\item{Cerdonio M, Morante S, Torresani D, Vitale S, de
Young A, Noble RW. Reexamination of the evidence
for paramagnetism in oxy- and carbonmonoxyhaemoglobin.
Proc Natl Acad Sci USA 1983; 82:102Ð103.}
\item{Kosaka H, Tyuma I, Imaizumi K. Mechanism of autocatalytic
oxidation of oxyhemoglobin by nitrite.
Biomed Biochim Acta 1983; 42:S144ÐS148.}
\item{Skibsted U, Hansen PE. $^1$H NMR spin echo spectroscopy
of human erythrocytes. Transformation of exogenous
compounds. NMR Biomed 1990; 3:248Ð258.}
\item{Hansen P, Skibsted, Rae CD, Kuchel PW. $^1$H NMR of
compounds with low water solubility in the presence of
erythrocytes: effects of emulsion phase separation. Eur
Biophys J 2001; 30:69Ð74.}
\item{Jackman L, Sternhell S. Applications of Nuclear Magnetic
Resonance Spectroscopy in Organic Chemistry.
Oxford, UK: Pergamon Press; 1969. p. 203.}
\item{Kirk K, Kuchel PW. The contributions of magnetic
susceptibility effects to transmembrane chemical shift
differences in the $^{31}P$ NMR spectra of oxygenated
erythrocyte suspensions. J Biol Chem 1989; 263:130Ð
134.}
\item{Kuchel PW, Chapman BE, Xu AS-L. Rates of anion
transfer across erythrocyte membranes measured with
NMR spectroscopy. In: Bamberg E, Passow H, eds. The
Band 3 Proteins: Anion Transporters, Binding Proteins
and Senescent Antigens. Amsterdam: Elsevier; 1992;
p.105Ð119.}
\item{Chu SC-K, Xu Y, Balschi JA, Springer CS. Bulk magnetic-
susceptibility shifts in NMR-studies of compartmentalized
samplesÑuse of paramagnetic reagents.
Magn Reson Med 1990; 13:239Ð262.}
\item{Lin WR, Mota de Freitas DM. Cl-35 NMR study of Cl 
distribution and transport in human red blood cell suspensions.
Magn Reson Chem 1996; 34:768Ð772.}
\end{enumerate}

\end{document}